\documentclass[twocolumn,aps,floatfix,superscriptaddress,longbibliography]{revtex4-2}
\usepackage{amsmath,amssymb,eucal,graphicx,float,epstopdf,xparse,color,bm}
\usepackage{epsfig,subfigure}

\usepackage[colorlinks=true, urlcolor=blue, anchorcolor=blue, citecolor=blue,filecolor=blue,linkcolor=blue,menucolor=blue]{hyperref}

\newcommand{\Hil}{\mathcal{H}}
\def\Xint#1{\mathchoice
	{\XXint\displaystyle\textstyle{#1}}%
	{\XXint\textstyle\scriptstyle{#1}}%
	{\XXint\scriptstyle\scriptscriptstyle{#1}}%
	{\XXint\scriptscriptstyle\scriptscriptstyle{#1}}%
	\!\int}
\def\XXint#1#2#3{{\setbox0=\hbox{$#1{#2#3}{\int}$}
		\vcenter{\hbox{$#2#3$}}\kern-.5\wd0}}
\def\dashint{\Xint-}

\begin{document}
\title{Expansion into the vacuum of stochastic gases with long-range interactions}

\author{P. L. Krapivsky}
\affiliation{Department of Physics, Boston University, Boston, Massachusetts 02215, USA}
\affiliation{Santa Fe Institute, Santa Fe, New Mexico 87501, USA}
\author{Kirone Mallick}
\affiliation{Institut de Physique Th\'{e}orique, CEA/Saclay, F-91191 Gif-sur-Yvette Cedex, France}

\begin{abstract}
We study the evolution of a system of many point particles initially concentrated in a small region in $d$ dimensions. Particles undergo overdamped motion caused by pairwise interactions through the long-ranged repulsive $r^{-s}$ potential; each particle is also subject to white Gaussian  noise. When $s<d$, the expansion is governed by non-local hydrodynamic equations. In the one-dimensional case, we deduce self-similar solutions for all $s\in (-2,1)$. The expansion of Coulomb gases remains well-defined in the infinite-particle limit: The density is spatially uniform and inversely proportional to time independent of the spatial dimension. 
\end{abstract}
  
\maketitle

\section{Introduction}

Many-particle systems with long-range interactions \cite{Ruffo09,Ruffo14} display unusual collective behaviors and fluctuations. Probing these behaviors requires an extension of the techniques (such as hydrodynamic limits and large deviations) initially developed studies of systems with short-range forces.

In the present work, we consider point particles in $\mathbb{R}^d$ undergoing independent Brownian motions and an overdamped motion caused by the repulsive Riesz potential
\begin{equation}
\label{V-Riesz}
V(\mathbf{r})= \frac{g}{sr^s} \, \hbox{ with } g > 0 \,.
\end{equation}
When the decay exponent $s$ is smaller than the dimension of space, $s<d$, the interaction is long-ranged, viz., it dominates diffusion for a sufficiently long time, even forever, if $s<0$. In the present work, we consider long-ranged Riesz potentials.  Brownian particles interacting via power-law repulsive forces appear in various physical systems in soft condensed matter such as Coulombic and dipolar fluids \cite{NOppenheimer}, liquid crystals, defects in membranes, Marangoni flows in interfacial science \cite{Marangoni95,Marangoni22,Marangoni24,Diamant2025}. Non-local hydrodynamic equations governing the evolution of such systems sometimes indicate finite-time blowups. Such equations provide toy models of blowup phenomena \cite{blowup96,blowup20}.

In the $s\to 0$ limit, the Riesz potential becomes logarithmic
\begin{equation}
\label{V-log}
V(\mathbf{r})= -g\ln r
\end{equation}
In one dimension, a gas of Brownian particles interacting via logarithmic potential was introduced by Dyson \cite{Dyson62} in the context of random matrices \cite{Dyson62,Forrester} where eigenvalues played the role of particles. In two dimensions, a gas of  Brownian particles interacting via logarithmic potential was introduced by Ginibre \cite{Ginibre65} also in the context of random matrices, with complex eigenvalues playing the role of particles \cite{Ginibre65,Forrester,Byun24}. In the Dyson and Ginibre gases, the particles undergo an overdamped motion caused by the logarithmic potential and are subject to Brownian noises. The Dyson and Ginibre gases suggest exploring stochastic Riesz gases with arbitrary long-ranged Riesz potential and in arbitrary dimensions. Deterministic Riesz gases with particles moving deterministically and interacting through Riesz potential is a more explored subject  \cite{Lewin22,Serfaty24}, but in this paper, we consider only {\em stochastic} Riesz gases.

Apart from logarithmic gases corresponding to $s=0$, several other values of the exponent $s$ correspond to interesting classical models. The $r^{-2}$ potential is the Calogero gas mostly studied in one dimension \cite{Calogero69,Alexios06}, albeit it makes sense in arbitrary dimension $d$; the Calogero gas is long-ranged when $d>2$. When $s=d-2$, the Riesz potential reduces to the Coulomb potential
\begin{equation}
\label{V-Coulomb-d}
V(\mathbf{r})= \frac{g}{(d-2)r^{d-2}} 
\end{equation}
In two dimensions, the Coulomb potential is logarithmic. 

The equation of motion of the position $\mathbf{x}_i(t)\in \mathbb{R}^d$ of the particle with label $i$ is 
\begin{equation}
\label{micro}
\frac{d\mathbf{x}_i}{dt} = g \sum_{j \neq i} \frac{\mathbf{x}_i -  \mathbf{x}_j}{|\mathbf{x}_i -  \mathbf{x}_j|^{2 +s}} + \boldsymbol{\eta}_i (t) 
\end{equation}
We emphasize that the system is overdamped and thus differs from Riesz gases of Newtonian particles \cite{Lewin22,Serfaty24}. Each particle performs an overdamped motion caused by interactions with other particles leading to the deterministic contribution to the velocity, the first term on the right-hand side of Eq.~\eqref{micro}. There is also a stochastic contribution caused by the white noise $ \boldsymbol{\eta}_i (t)$. The noises have vanishing average, are independent and  would cause each particle to diffuse with diffusion coefficient $D$ if it were alone, i.e., $\langle \boldsymbol{\eta}_i (t) \boldsymbol{\eta}_j (t) \rangle = 2 D \delta_{ij} \delta(t - t').$  The noises represent the interactions with the background thermal bath and the diffusion constant $D$ is finite, proportional to the temperature. However, according to the context, the noise terms can have different physical  interpretation: for example, in the astrophysical literature, external noise terms are introduced to model collisional stellar dynamics \cite{Chandrasekhar,Kandrup1,Kandrup2,Kandrup01}, and the noise strength $D$ is typically proportional to $1/N$.

We shall always assume that the number of particles is large, $N\gg 1$. In this situation, the collective behavior of the system is well described by a continuous (hydrodynamic) framework. The chief difference with standard hydrodynamic equations is that the governing equations are non-local. The influence of noise is limited or negligible throughout the evolution, as will become clear from our analysis.

For a  logarithmic potential, the interaction strength $g$ and the diffusion coefficient $D$ have the same dimension $[\text{length}]^2/[\text{time}]$. Their relative strength is quantified by the (dimensionless) ratio
\begin{equation}
  \beta =  \frac{g}{D} 
\label{def:beta}
\end{equation}
playing the role of an inverse temperature \cite{Lewin22}. For the one-dimensional Dyson gas \cite{Dyson62} introduced in the context of random matrices, the exceptional values $\beta=1,2,4$ correspond to canonical ensembles (symmetric, hermitian, symplectic) of random matrices \cite{Dyson62, Forrester,Vivo18}.

Our goal is to analyze the expansion into the vacuum of a large number of particles initially concentrated in a small spatial region. We rely on a continuous framework, namely on non-local hydrodynamic equations accounting for long-ranged interactions and ignoring diffusion. In one dimension, the expansion of a long-ranged Riesz gas with $s\in (-2,1)$ can be solved exactly. In higher dimensions, $ d \ge 2$, the  Coulomb gases are readily tractable thanks to the Newton-Gauss theorem. For the non-Coulombic Riesz gases, $s\ne d-2$, in $d\geq 2$ dimensions, the functional form of the exact solution is the straightforward generalization of the one-dimensional solution, but the derivation is much more challenging. Further remarks about this case are given in Sec.~\ref{sec:remarks}.

We now present density profiles for a few Riesz gases with long-range potential ($s<d$). The common feature is that these profiles have compact support. For the one-dimensional Riesz gas with $s\in (-2,1)$, the density 
\begin{equation}
\label{rho-sol:R}
\rho(x,t) = \frac{B_s}{gt}\,(L^2-x^2)^\frac{s+1}{2}\,, \quad B_s = \frac{\cos(\pi s/2)}{\pi(s+1)(s+2)}
\end{equation}
is non-vanishing on the interval $[-L(t), L(t)]$ with 
\begin{equation}
\label{L:R}
L(t) = \left[\frac{\sqrt{\pi}\,(2+s)^2\,\Gamma\big(1+\frac{s}{2}\big)}{\cos\!\big(\frac{\pi s}{2}\big)\,\Gamma\big(\frac{1+s}{2}\big)}\,Ngt\right]^\frac{1}{s+2}
\end{equation}
For the Dyson gas ($s=0$), the density profile \eqref{rho-sol:R} is the celebrated Wigner semi-circle (expanding in the present case without confining potential). 

For Coulomb gases in arbitrary dimension, $s=d-2$, the radius of the expanding ball is 
\begin{equation}
\label{R:Coulomb}
R=(d N gt)^{1/d}
\end{equation}
The density profiles are uniform inside the expanding ball of radius $R(t)$ and vanish outside the ball. The density of stochastic Coulomb gases inside the expanding ball is remarkably universal, viz., independent on $N$ and inversely proportional to time in arbitrary spatial dimension: 
\begin{equation}
\label{Friedmann}
\rho(r,t) = (\Omega_d g t)^{-1}
\end{equation}
where $\Omega_d$ is the volume of the unit sphere $\mathbb{S}^{d-1}$. 

Intriguingly,  the expansion of Coulomb gases remains well-defined when $N=\infty$. For non-Coulombic stochastic Riesz gases with infinitely many particles, the problem is ill-defined: the density vanishes when $s<d-2$ and becomes infinite when $s\in (d-2,d)$.

The outline of this work is as follows. In Sec.~\ref{sec:1d}, we solve the expansion problem for one-dimensional long-ranged Riesz gases with $s\in (-2,1)$. In  Sec.~\ref{sec:CoulombHighd}, we analyze the expansion problem for the Coulomb gases in arbitrary dimensions. We provide a detailed derivation in two spatial dimensions, i.e., for the Ginibre gas (Sec.~\ref{sec:2d}), and in three dimensions, i.e., for the classical Coulomb gas (Sec.~\ref{sec:d}). In Sec.~\ref{sec:remarks}, we discuss open problems and mention amusing analogies between our results and formulas arising in Newtonian cosmology.

\section{One dimension}
\label{sec:1d}

At the coarse grained level, the behavior of the large number of particles evolving according to Eqs.~\eqref{micro} is described by continuum field theory for a single scalar field, the density $\rho(x,t)$, satisfying the continuity equation which in one dimension reads
\begin{equation}
\label{continuity}
\partial_t \rho + \partial_x J = 0
\end{equation}
The local current $J(x,t)$ depends on the density and the inter-particle potential. The conservation equation \eqref{continuity} ensures that the total number of particles remains constant throughout the evolution:
\begin{equation}
\label{mass}
\int_{-\infty}^\infty dx\,\rho(x,t)=N
\end{equation}
at all $t\geq 0$. 

To close the continuity equation \eqref{continuity}, we must express $J$ through the density and the interaction potential. The current contains the standard diffusion term, $-D\partial_x \rho$, resulting from the entropic contribution at microscopic level. A general long-range interaction potential $V$ gives rise to a non-local deterministic contribution 
\begin{equation}
  J_V(x,t) =  \rho(x,t)\int_{-\infty}^\infty dy\,\nabla V(x -y) {\rho(y,t)} 
  \label{def:Jv}
\end{equation}
to the current. There is also a stochastic component to the current that can be written as $\sqrt{ 2 \rho}\, \eta$, where the noise $\eta(x,t)$ is a Gaussian noise, white in space and time. The amplitude $\sqrt{ 2 \rho}$ of this noise originates from the Brownian character of the particles, see \cite{Spohn91}. Thus, the total current $J(x,t)$ can be written as
\begin{equation}
  J = -D\partial_x \rho + J_V + \sqrt{ 2 \rho}\, \eta
  \label{fullJ}
\end{equation}
A formal derivation of \eqref{fullJ} by taking the continuous limit of the discrete set of equations can be obtained by following the strategy developed by Kawasaki and Dean \cite{kawasaki1994,dean96} (see \cite{touzo2023interacting,RahulDyson} for recent expositions).

We consider one-dimensional stochastic Riesz gases with $s\in (-2,1)$. If $s>1$, the interactions are effectively short-ranged (see \cite{Kundu19,RahulRiesz}), the governing hydrodynamic equations are local, and the emerging behaviors are more standard. The usual justification of the lower bound is that the original appearance of the Dyson gas in the context of random matrices involves the harmonic confining $r^2$ potential, and one does not want the pairwise interactions to overwhelm the confining potential. Even without confining potential, the systems with repulsive $r^{-s}$ interactions are very peculiar when $s<-2$. Already for two particles, the inter-particle separation $\ell(t)$ satisfies $\frac{d\ell}{dt}=g\ell^{-s-1}$, so $\ell$ diverges in a finite time if $s<-2$. (This happens in arbitrary spatial dimension; taking diffusion into account does not prevent the finite-time blow-up \cite{Meerson25}.) These arguments explain why we study Riesz gases with exponents $s\in (-2,1)$, and generally $s\in (-2,d)$ in $d$ dimensions. 

In the present work, we focus on the typical gas expansion. We see that the influence of diffusion on the expansion of Riesz gases with $N\gg 1$ particles depends on whether $s<0$ or $s>0$. When $-2<s<0$, the expansion is super-diffusive, namely $L(t)\sim (Ngt)^\frac{1}{s+2}$, cf. Eq.~\eqref{L:R}, and diffusion is effectively irrelevant. Whe $0<s<1$, the expansion is sub-diffusive. The diffusion length $\sqrt{Dt}$ catches $L(t)$ at a crossover time $t_*$ estimated from $ \sqrt{ Dt_*}  \sim (Ng t_*)^\frac{1}{s+2}$. Thus, the crossover time and size $L_*=L(t_*)$ of the occupied region are
\begin{equation}
\label{cross-time}
t_* \sim D^{-1} \left( \frac{ Ng}{D}\right)^{2/s}\,,\quad L_* \sim \left( \frac{ Ng}{D}\right)^{1/s}
\end{equation}
This crossover time diverges when $N \to \infty$, as expected. For $ t \ll t_*$, the profile evolves as \eqref{rho-sol:R} whereas for $ t \gg t_*$, the ordinary diffusion prevails.

The hydrodynamic description predicts that the occupied region is compact, $[-L(t),L(t)]$ if we ignore diffusion. For Riesz gases with $s\in (0,1)$, diffusion cannot be ignored when $t\gtrsim t_*$. For Riesz gases with $s\in (-2,0)$, the continuum description is inapplicable outside the $[-L(t),L(t)]$ region. The analogy with random matrices suggests the position of the right-most [resp. left-most] particle is close to $L(t)$ [resp. $-L(t)$]. Studying the distributions of the positions of the right-most and left-most particles requires different techniques, so we shall merely present a few estimates but mostly rely on the hydrodynamic description accounting only the non-local deterministic contribution $J_V$ to the current \eqref{fullJ}. 

\subsection{Dyson gas}

The Dyson gas separates the $s\in (-2,0)$ range where diffusion is irrelevant and $s\in (0,1)$ range where diffusion is relevant when $t\gtrsim t_*$. Let us keep the diffusive contribution in the marginal case of the Dyson gas. Specializing \eqref{def:Jv} to logarithmic interactions, we obtain the component of the current in the Dyson gas caused by interactions. One gets (see also \cite{touzo2023interacting,RahulDyson}) 
\begin{equation}
\label{current}
J_V(x,t)=g \rho(x,t)\dashint_{-\infty}^\infty dy\,\frac{\rho(y,t)}{x-y}
\end{equation}
where $\dashint$ denotes the Cauchy principal value 
\begin{equation*}
\dashint_{-\infty}^\infty dy\, \frac{\rho(y,t)}{x-y}=\lim_{\delta \to 0}\int_{|x-y|> \delta}
   dy\, \frac{\rho(y,t)}{x-y}\,.
\end{equation*}
Recalling the definition of the Hilbert transform
\begin{equation}
   \Hil[\rho](x,t) = \frac{1}{\pi}\dashint_{-\infty}^\infty dy\, \frac{\rho(y,t)}{x-y}
   \label{def:Hilb}
\end{equation}
and adding the diffusion term we arrive at 
\begin{equation}
J = -D\partial_x \rho + \pi g \rho \Hil[\rho]
\end{equation}
Combining this current with \eqref{continuity} we obtain 
\begin{equation}
\label{Dyson-Hilb}
\partial_t \rho + \partial_x\!\left(\pi g \rho \Hil[\rho]\right)
 -D \partial_{xx} \rho  = 0
\end{equation}

The expansion of the Dyson gas into the vacuum begins with particles concentrated in a tiny spatial region. We thus postulate that all particles are initially at the origin: 
\begin{equation}
\label{IC:origin}
\rho(x,0)=N\delta(x)
\end{equation}

The initial-value problem \eqref{Dyson-Hilb}--\eqref{IC:origin} is invariant under the one-parameter group of transformations
\begin{equation}
\label{group-a}
\rho\to a^{-1} \rho, \qquad x\to a x, \qquad t\to a^2 t.
\end{equation}
This implies that $\sqrt{t} \rho(x,t)$ depends on the single variable $x/\sqrt{t}$. A more carefully chosen self-similar form 
\begin{equation}
\label{scaling:D}
\rho(x,t) = \frac{N}{\sqrt{Ng t}}\,F(X)\quad \text{with} \quad X = \frac{x}{\sqrt{N g t}}
\end{equation}
has an advantage that both $N$ and $g$ disappear from the leading terms in the governing equation for the scaled density $F(X)$ and from the conservation law. Indeed, the conservation law \eqref{mass} turns into the conservation law
\begin{equation}
\label{F:t=0}
\int_{-\infty}^\infty dX\, F(X) = 1
\end{equation}
Substituting the scaling ansatz \eqref{scaling:D} into \eqref{Dyson-Hilb} we find that the scaled density satisfies
\begin{equation*}
\frac{d(XF)}{dX} + \frac{2}{\beta N} \,\frac{d^2F}{dX^2} = 2 \frac{d}{dX} \left[F(X) \dashint_{-\infty}^\infty dY\,\frac{F(Y)}{X-Y}\right]
\end{equation*}
which we integrate to find
\begin{equation}
\label{F-eq-long}
X F  + \frac{2}{\beta N} \,\frac{dF}{dX} = 2 F(X) \dashint_{-\infty}^\infty dY\,\frac{F(Y)}{X-Y}
\end{equation}

The non-local Eq.~\eqref{Dyson-Hilb} has been studied in the realm of the fluid mechanical problem of Marangoni spreading, and several exact solutions have been found \cite{Marangoni22,Marangoni24}. One can extract an exact solution of the non-linear integro-differential equation \eqref{F-eq-long} from \cite{Marangoni22,Marangoni24}. This solution is very cumbersome, so we do not present it. Since we are interested  in the behavior of a large number particles, we first neglect the $1/N$ term in Eq.~\eqref{F-eq-long}, and then briefly outline an asymptotic analysis applicable when $N\gg 1$. When $N=\infty$,  i.e., neglecting the diffusion term in the original problem, we  arrive at
\begin{equation}
\label{F-eq-short}
F(X)\left[X - 2 \dashint_{-\infty}^\infty dY\,\frac{F(Y)}{X-Y}\right]=0
\end{equation}
We seek a symmetric solution, $F(X)=F(-X)$, vanishing in the $|X|\to\infty$ limits. Such solution of Eq.~\eqref{F-eq-short} has a compact support: $F=0$ for $|X|>R$; when $|X|<R$, the scaled density satisfies 
\begin{equation}
\label{F-eq}
X = 2\dashint_{-R}^R dY\,\frac{F(Y)}{X-Y}\,.
\end{equation}
The solution of the linear integral equation \eqref{F-eq} is well-known, $F=(2\pi)^{-1}\sqrt{R^2-X^2}$.  [See \cite{Estrada,Polyanin,Carrillo17} for the descriptions of the methods of solving Eq.~\eqref{F-eq} and similar linear singular integral equations.] Using the conservation law \eqref{F:t=0}, we fix $R=2$. Summarizing 
\begin{equation}
\label{F-sol}
F = 
\begin{cases}
\frac{1}{2\pi} \sqrt{4-X^2}  & |X|<2\\
0                                       & |X|>2
\end{cases}
\end{equation}
In the original variables, the density profile is the Wigner semi-circle 
\begin{equation}
\label{Wigner}
\rho(x,t) = \frac{1}{2\pi g t}\,\sqrt{4Ngt-x^2}
\end{equation}
as for the equilibrium Dyson gas in a harmonic confining potential. The difference is that the span now diffusively expands with time: $|x|\leq 2\sqrt{Ngt}$.

One can employ asymptotic techniques \cite{Bender} to obtain more precise solution of Eq.~\eqref{F-eq-long} inside the occupied $|X|<2$ region. One seeks the solution as an expansion in a small parameter $(\beta N)^{-1}\ll 1$:
\begin{equation}
\label{FX:exp}
F(X)=\frac{1}{2\pi}\, \sqrt{4-X^2}+\frac{1}{\beta N}\,F_1(X) + \ldots 
\end{equation}
Sustituting \eqref{FX:exp} into Eq.~\eqref{F-eq-long} we find that the leading correction satisfies 
\begin{equation}
\label{F1-eq}
\dashint_{-2}^2 dY\,\frac{F_1(Y)}{X-Y}=-\frac{X}{4-X^2}
\end{equation}
We solve the linear singular integral equation \eqref{F1-eq} using the methods described in \cite{Estrada,Polyanin,Carrillo17} and obtain 
\begin{equation}
\label{F1:sol}
F_1(X)=-\frac{1}{\pi^2\sqrt{4-X^2}}\left[4+X\,\ln\frac{2-X}{2+X}\right]
\end{equation}
The leading and the sub-leading terms in the expansion \eqref{FX:exp} become comparable when $2-|X|\sim N^{-1}$. Therefore,  Eq.~\eqref{FX:exp} gives an outer expansion, which should be matched with inner expansions valid in the internal $2-|X|\sim N^{-1}$ layers. We do not perform such an analysis since, inside the internal layers, the deterministic framework is no longer justified. (Relying on the deterministic continuum framework for $|X| > 2$ is even more erroneous.) Indeed, using \eqref{Wigner}, one estimates the positions $x_\pm(t)$ of the right-most and left-most particles:
\begin{equation}
\label{extreme:D}
\frac{x_\pm(t)}{L(t)} = \pm 1+N^{-\frac{2}{3}}\,\xi_\pm(\beta)
\end{equation}
with $L(t)=2\sqrt{Ngt}$. The random variables $\xi_\pm(\beta)$ are expected to be asymptotically stationary with Tracy-Widom type distributions dependent on $\beta=g/D$. (The Tracy-Widom distributions appearing in the context of random matrices are explicitly known for $\beta=1,2,4$ characterizing ensembles of random matrices \cite{Forrester,Majumdar14,Vivo18}. In our case, $\beta>0$ is arbitrary.) 

\subsection{General case}

\begin{figure}
\centering
\includegraphics[width=7.7cm]{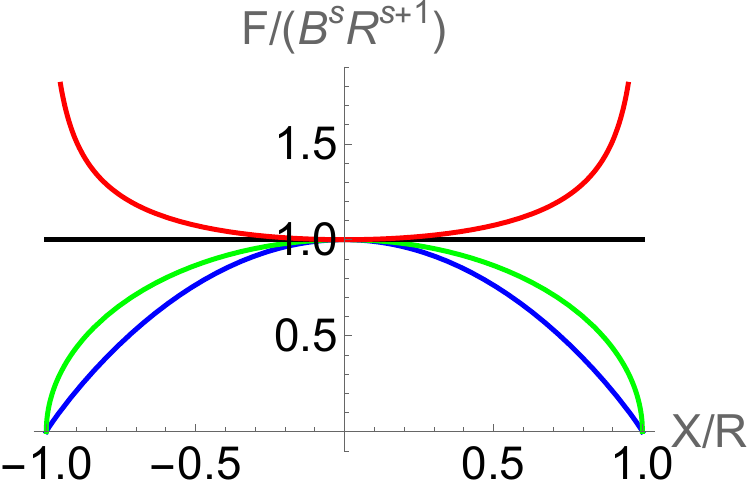}
\caption{The spatial dependence of the density re-scaled to be equal to unity at the origin: $F/(B_s R^{s+1})$, with $F$ given by Eq.~\eqref{F-sol:R}, versus $X/R$. The density vanishes when $|X|>R$. The shown curves  (top to bottom) correspond to the Riesz exponent $s=-3/2, -1, 0, 7/8$. The density is flat for the one-dimensional Coulomb gas ($s=-1$). The semi-circle represents the density of the Dyson gas ($s=0$). }
\label{Fig:density-R}
\end{figure}

We now consider general one-dimensional Riesz gases with long-ranged interactions, $-2<s<1$. We ignore diffusion and comment below when this approximation is justifiable. The contribution \eqref{def:Jv} to the current due to interactions becomes (see also \cite{RahulRiesz})
\begin{equation}
\label{current:R}
J(x,t)=g \rho(x,t)\dashint_{-\infty}^\infty dy\,\frac{x-y}{|x-y|^{s+2}}\, \rho(y,t)
\end{equation}
Combining the  continuity equation \eqref{continuity} with \eqref{current:R} leads to the following integro-differential equation 
\begin{equation}
\label{Riesz:1d}
\partial_t \rho + \partial_x\!\left[g \rho \dashint_{-\infty}^\infty dy\,\frac{x-y}{|x-y|^{s+2}}\, \rho(y,t)\right] = 0
\end{equation}
for the density profile. The governing equation \eqref{Riesz:1d} and the conservation law  \eqref{mass} are invariant under the one-parameter group of transformations
\begin{equation}
\label{group:R}
\rho\to a^{-\frac{1}{s+2}}  \rho, \qquad x\to a^{\frac{1}{s+2}} x, \qquad t\to a t.
\end{equation}
Using the self-similar form 
\begin{equation}
\label{scaling}
\rho(x,t) = \frac{N}{(Ng t)^\frac{1}{s+2}}\,F(X)\quad \text{with} \quad X = \frac{x}{(Ng t)^\frac{1}{s+2}}
\end{equation}
we find that the scaled density satisfies 
\begin{equation}
\label{F:R}
\frac{F + X\,\frac{dF}{dX}}{s+2} = \frac{d}{dX}\left[F\dashint_{-\infty}^\infty dY\,\frac{X-Y}{|X-Y|^{s+2}}\,F(Y)\right]
\end{equation}
Integrating Eq.~\eqref{F:R} we arrive at the integral equation
\begin{equation}
\label{F1:eq}
\frac{X}{s+2} = \dashint_{-R}^R dY\,\frac{X-Y}{|X-Y|^{s+2}}\,F(Y)
\end{equation}
inside the $|X|<R$ interval where $F(X)>0$; the density vanishes outside that interval. Integral equations such as  \eqref{F1:eq} arise in many problems, see, e.g., \cite{Carrillo17,Kundu19}. The solution has again a compact support and appears in textbooks  \cite{Estrada,Polyanin}. For  $|X|<R$, we have 
\begin{equation}
\label{F-sol:R}
F = B_s (R^2-X^2)^\frac{s+1}{2}\,, \quad B_s = \frac{\cos(\pi s/2)}{\pi(s+1)(s+2)}
\end{equation}
The normalization condition is again given by  \eqref{F:t=0} which we combine with \eqref{F-sol:R} to yield 
\begin{equation}
\label{R:R}
R = \left[\frac{\sqrt{\pi}\,(2+s)^2\,\Gamma\big(1+\frac{s}{2}\big)}{\cos\!\big(\frac{\pi s}{2}\big)\,\Gamma\big(\frac{1+s}{2}\big)}\right]^\frac{1}{s+2}
\end{equation}
The density is non-vanishing when $x\in [-L(t), L(t)]$ with $L(t) = R(Ng t)^\frac{1}{s+2}$ as was announced in Eq.~\eqref{L:R}. The density profile \eqref{F-sol:R} reduces to the announced result \eqref{rho-sol:R} in the original variables. In Fig.~\ref{Fig:density-R}, we show the density profiles for $s=-1,0,7/8$.
 
We relied on Eq.~\eqref{Riesz:1d} ignoring diffusion. If $0<s<1$, the diffusion is irrelevant when $t\ll t_*$ with crossover time given by \eqref{cross-time}. In the  $-2 < s < 0$ range, the expansion is super-diffusive and diffusion is essentially irrelevant. Using \eqref{rho-sol:R}--\eqref{L:R}, one estimates the positions of the right-most and left-most particles and arrives to the result similar to \eqref{extreme:D}, with $N^{-2/3}$ replaced by $N^{-\frac{2}{s+3}}$.

\section{Coulomb gases in higher dimensions}
\label{sec:CoulombHighd}

\subsection{Two dimensions (Ginibre gas)}
\label{sec:2d}

The expansion is isotropic. Hence $\rho(\mathbf{r},t)=\rho(r,t)$ and the continuity equation reduces to
\begin{equation}
\label{HD-Ginibre}
\partial_t \rho + r^{-1}\partial_r (rJ) = 0
\end{equation}
A simple expression for the current  
\begin{equation}
\label{current:2d}
J(r,t)=g \rho(r,t) r^{-1}\int_0^r dr'\,2\pi r' \rho(r',t)
\end{equation}
reflects the rotational symmetry and the Coulomb nature of the logarithmic potential in two dimensions: the force at $r$ is the same as if the entire charge of the disk of radius $r$ was at the origin (Gauss theorem). Introducing an auxiliary variable  
\begin{equation}
\label{F-2d:def} 
F(r,t) = 2\pi r \rho(r,t)
\end{equation}
we re-write \eqref{HD-Ginibre}--\eqref{current:2d} as 
\begin{equation}
\label{F-Ginibre}
\partial_t F(r,t) + g\partial_r\!\left[r^{-1}F(r,t)\int_0^r dr'\,F(r',t)\right] = 0
\end{equation}
The conservation law $\int d\mathbf{r}\, \rho(\mathbf{r},t)=N$ simplifies to
\begin{equation}
\label{mass:F}
\int_0^\infty dr\,F(r,t)=N
\end{equation}

Equations \eqref{F-Ginibre}--\eqref{mass:F} are invariant under the one-parameter group of transformations
\begin{equation}
\label{group:F}
F\to a^{-1} F, \qquad r\to a r, \qquad t\to a^2 t
\end{equation}
implying that the solution has a self-similar form
\begin{equation}
\label{scaling:F}
F(r,t) = \sqrt{\frac{N}{g t}}\,\Phi(\xi)\quad \text{with} \quad \xi = \frac{r}{\sqrt{N g t}}
\end{equation}
chosen to ensure that both $N$ and $g$ disappear from the governing equation and the conservation law for $\Phi(\xi)$.
Substituting the scaling ansatz \eqref{scaling:F} into \eqref{F-Ginibre} yields 
\begin{equation}
\label{Phi-eq-long}
\Phi + \xi\,\frac{d\Phi}{d\xi} = 2 \frac{d}{d\xi}\!\left[\frac{\Phi \Psi}{\xi}\right]
\end{equation}
where 
\begin{equation}
\label{Psi-def}
\Psi(\xi) = \int_0^\xi d\xi'\,\Phi(\xi')
\end{equation}

Integrating \eqref{Phi-eq-long} gives
\begin{equation}
\label{Phi-eq}
\Phi\left(\xi- \frac{2\Psi}{\xi}\right)=0
\end{equation}
The solution reads
\begin{equation}
\label{Phi-sol}
\Phi = 
\begin{cases}
\xi  & \xi\leq \xi_0\\
0    & \xi >\xi_0
\end{cases}
\end{equation}
Substituting this solution into the conservation law 
\begin{equation}
\label{mass:2d}
\int_0^\infty d\xi\,\Phi(\xi)=1
\end{equation}
we fix $\xi_0=\sqrt{2}$. Thus, at any time the density is uniform inside the growing disk 
\begin{equation}
\label{rho-sol}
\rho(r,t) = 
\begin{cases}
(2\pi g t)^{-1} & r\leq \sqrt{2Ngt} \\
0                    & r > \sqrt{2Ngt}
\end{cases}
\end{equation}

The radius of the disk grows diffusively as for the one-dimensional Dyson gas; the extra $N$ factor ensures again that diffusion plays a minor role. 
The diffusive growth of the ball indicates that the scaling ansatz \eqref{scaling:F} remains applicable even when we take diffusion into account. The auxiliary function \eqref{Psi-def} satisfies a neat ordinary differential equation 
\begin{equation}
\label{Psi-eq}
\frac{1}{2\beta N}\,\eta \Psi'' = \Psi'(\Psi-\eta)
\end{equation}
where $\eta=\xi^2/2$ and prime denotes a derivative with respect to $\eta$. This equation is simpler than the integro-differential equation \eqref{F-eq-long} describing the Dyson gas. Still, Eq.~\eqref{Psi-eq} is non-linear and analytically intractable.

\subsection{Arbitrary dimension}
\label{sec:d}

We begin with physically most relevant three-dimensional case. The expansion is spherically symmetric, and hence the continuity equation reads
\begin{equation}
\label{cont:3d}
\partial_t \rho + r^{-2}\partial_r (r^2J) = 0
\end{equation}
For the Coulomb gas, we have  $V(\mathbf{r})= g/r$ and
the current is given by
\begin{equation}
\label{current:3d}
J(r,t)=g \rho(r,t) r^{-2}\int_0^r dr'\,4\pi r'^2 \rho(r',t)
\end{equation}
This expression reflects rotational symmetry and  we have used again the Coulomb nature of the potential: the force at $r$ is the same as if the entire mass of the ball of radius $r$ was at the origin. Using 
\begin{equation}
\label{F-3d:def} 
F(r,t) = 4\pi r^2 \rho(r,t)
\end{equation}
we re-write \eqref{cont:3d}--\eqref{current:3d} as 
\begin{equation}
\label{F-3d}
\partial_t F(r,t) + g\partial_r\!\left[r^{-2}F(r,t)\int_0^r dr'\,F(r',t)\right] = 0
\end{equation}
The conservation law has the same form \eqref{mass:2d} as in two dimensions. Equations \eqref{F-3d} and \eqref{mass:2d} are invariant under the one-parameter group of transformations
\begin{equation}
\label{group:F3}
F\to a^{-1} F, \qquad r\to a r, \qquad t\to a^3 t
\end{equation}
implying that the solution has a self-similar form
\begin{equation}
\label{scaling:F-3d}
F(r,t) = \frac{N}{\sqrt[3]{Ngt}}\,\Phi(\xi)\quad \text{with} \quad \xi = \frac{r}{\sqrt[3]{Ngt}}
\end{equation}
Substituting the scaling ansatz \eqref{scaling:F-3d} into \eqref{F-3d} yields 
\begin{equation}
\label{Phi-3d-long}
\Phi + \xi\,\frac{d\Phi}{d\xi} = 3 \frac{d}{d\xi}\!\left[\frac{\Phi \Psi}{\xi^2}\right]
\end{equation}
which is integrated to give
\begin{equation}
\label{Phi-3d}
\Phi\left(\xi- \frac{3\Psi}{\xi^2}\right)=0
\end{equation}
The solution reads 
\begin{equation}
\label{Phi-sol-3}
\Phi = 
\begin{cases}
\xi^2  & \xi\leq \xi_0\\
0        & \xi >\xi_0
\end{cases}
\end{equation}
Substituting this solution into the conservation law \eqref{mass:2d} we fix $\xi_0=\sqrt[3]{3}$. Thus at any time the density is uniform inside the growing disk 
\begin{equation}
\label{rho-3d}
\rho(r,t) = 
\begin{cases}
(4\pi g t)^{-1} & r\leq \sqrt[3]{3Ngt} \\
0                    & r > \sqrt[3]{3Ngt}
\end{cases}
\end{equation}
The radius of the ball grows sub-diffusively, so the above results are valid up to a crossover time $t_*\sim (Ng)^2/D^3$.

The above calculations bear some resemblance to the well-studied problem of Coulomb explosion, an important phenomenon in laser-matter interaction, in which a  nanoscale cluster, having all its electrons swept away  by an intense laser pulse, undergoes an explosive dynamics with some self-similar profiles\cite{Kaplan03,Kovalev,Grech11}. In recent work \cite{Zandi2022}, stochastic forces due to collisions of particles and random motion are incorporated by adding a pressure term to the hydrodynamic equations of motion. However, the model we  study  here  has  some important  differences with Coulomb explosion: we consider repulsive particles with overdamped dynamics (inertial effects are neglected);  the initial state has no internal structure (it is a pure Dirac function);  the approach used is based on stochastic density functional theory (Dean-Kawasaki equation) rather than on hydrodynamics; and finally, in the calculations above, diffusive effects have been neglected for times  $t \ll t_*$.

Extending to arbitrary spatial dimension we again obtain a uniform density profile 
\begin{equation}
\label{rho-d}
\rho(r,t) = 
\begin{cases}
(\Omega_d g t)^{-1} & r\leq (dNgt)^{1/d} \\
0                               & r > (dNgt)^{1/d}
\end{cases}
\end{equation}
where $\Omega_d = 2 \pi^{d/2}/\Gamma(d/2)$ is the volume of the unit sphere $\mathbb{S}^{d-1}$. Therefore the universal $t^{-1}$ decay occurs inside a ball with radius growing as $t^{1/d}$ and only the amplitude is dimension dependent. 

Consider now the expansion into the vacuum of an infinite free gas with infinitely many particles, $N=\infty$. Stunningly, the Coulomb gas is the only Riesz gas with a well-defined behavior in this limit. This claim is easy to prove for the one-dimensional Riesz gas solved in Sec.~\ref{sec:1d}. For instance, the density at the origin in the system with a large but finite number of particles 
\begin{equation}
\label{origin}
\rho(0,t) = (NC_s)^\frac{s+1}{s+2}\, [g(s+2) t]^{-\frac{1}{s+2}}
\end{equation}
diverges in the $N\to\infty$ limit if $s>-1$ and vanishes if $s<-1$. A  well-defined finite answer emerges only when $s=-1$, i.e., for the one-dimensional Coulomb potential. More generally, for the Coulomb gas in arbitrary dimension, the density is well-defined, uniform in the entire infinite space $\mathbb{R}^d$, and decays inversely proportional to time when the number of particles is infinite, $\rho(r,t) = (\Omega_d g t)^{-1}$.

\section{Concluding remarks}
\label{sec:remarks}

We have analyzed the expansion of the stochastic Riesz gases with long-ranged interactions in one dimension and of the Coulomb gases in an arbitrary dimension. Stochastic Coulomb gases remain well-defined in the infinite-particle limit, $N=\infty$. The emerging spatially uniform solution \eqref{Friedmann} bears an amusing resemblance with the uniform decaying density in the expanding Universe. The gravitational (Newton) potential is Coulombic yet attractive, but the motion is governed by Newton's laws (not overdamped). In the Universe filled with ``dust" (zero pressure limit), the density decays as $t^{-2}$ in Newtonian cosmology and also in Einsteinian cosmology in the flat Universe \cite{Liddle, Mukhanov}. The $t^{-2}$ decay is independent of the spatial dimension and the $t^{-1}$ decay \eqref{Friedmann} in our overdamped  `Universe' is also universal. The Hubble law for the velocity, $\mathbf{v}=H(t)\mathbf{r}$ with $H=\frac{2}{d} t^{-1}$, is replaced by a Hubble law for the current: $\mathbf{J}=H(t)\mathbf{r}$ with $H=(d \Omega_d g)^{-1} t^{-2}$. The underlying reason for these striking similarities is the peculiar nature of the Coulomb potential. 

We have not been able to analyze the expansion of the Riesz gas in $d >1$ dimensions, and we now explain the technical challenge. The governing equation  is given by
\begin{equation}
\label{cont:d-gen}
\partial_t \rho + \nabla\cdot \mathbf{J} = 0
\end{equation}
with
\begin{equation}
\label{current:R-d}
 \mathbf{J}(\mathbf{r},t)=g \rho(\mathbf{r},t)\int_{B_R} d\mathbf{y}\,\frac{\mathbf{r}-\mathbf{y}}{|\mathbf{r}-\mathbf{y}|^{s+2}}\, \rho(\mathbf{y},t)
\end{equation}
This equation can be interpreted as a non-local porous medium equation. The existence of explicit self-similar solutions and weak solutions in a more general class of non-local porous medium equations have recently been investigated \cite{Biler11,Biler2}.

We expect that the density $\rho(\mathbf{r},t)$ vanishes outside the ball $B_R$ of radius $R$, so we integrate over this ball, i.e., $|\mathbf{y}|<R$, in \eqref{current:R-d}. The rotational symmetry implies $\rho(\mathbf{r},t)= \rho(r,t)$ and $\mathbf{J}(\mathbf{r},t)=\frac{\mathbf{r}}{r}\,J(r,t)$, so Eq.~\eqref{cont:d-gen} simplifies to 
\begin{equation}
\label{cont:d}
\partial_t \rho + r^{-(d-1)}\partial_r (r^{d-1}J) = 0
\end{equation}
The functional $J[\rho]$ involves a $d-$fold integral of $\rho$, which can not be simplified further because for non-Coulombic gases where we cannot rely on the Newton-Gauss theorem. Note that  similar difficulties  appear in studies of `gravitational' collapse for attractive $1/r^s$  potentials with arbitrary $s$  \cite{Ispolatov,DiCintio1,DiCintio2,Chavanis1} (see also \cite{Miller,Tremaine,Mann} for further discussions about one-dimensional systems); in the investigation of the statistical properties of the force field for a Poisson distribution of particles \cite{Joyce1,Joyce2,Joyce3};  in developing a kinetic theory for long-range interacting systems with an arbitrary potential of interaction \cite{Chavanis2006,Chavanis2010,Chavanis2012,Fouvry2020}; etc.

The self-similar form \eqref{scaling} generalizes to 
\begin{equation}
\label{scaling:R-d}
\rho(r,t) = \frac{N}{(Ng t)^\frac{d}{s+2}}\,F(\xi)\quad \text{with} \quad \xi = \frac{r}{(Ng t)^\frac{1}{s+2}}
\end{equation}
The Riesz exponent varies in the $s\in (-2, d)$ range. The expected scaled density reducing to \eqref{F-sol:R} in one dimension and compatible with the uniform density for the Coulomb gas in arbitrary dimension is 
\begin{equation}
\label{FR:d}
F(\xi)=
\begin{cases}
B(\xi_0^2-\xi^2)^\frac{s+2-d}{2}   &\xi<\xi_0\\
0                                                  & \xi>\xi_0
\end{cases}
\end{equation}
Substituting the density \eqref{scaling:R-d}--\eqref{FR:d} into the conservation law $\int d{\bf r} \rho(r,t) = N$ gives a relation between $B$ and $\xi_0$:
\begin{equation}
B \xi_0^{s+2} = \frac{\Gamma\big(2+\frac{s}{2}\big)}{\pi^\frac{d}{2}\,\Gamma\big(2+\frac{s-d}{2}\big)} 
\end{equation}
The derivation of \eqref{scaling:R-d}--\eqref{FR:d} and fixing the ampltude $B$ and the re-scaled radius $\xi_0$ is rather involved, so we refer to \cite{Biler11,Biler2} for the analysis of the fundamental solutions of a class of non-local porous medium equations, \cite{Marangoni22} for the solution in the two-dimensional Riesz gas with $s=1$, and \cite{Vazquez} for review.

Stepping away from the unconstrained hydrodynamic behaviors, we mention fluctuations. For one-dimensional Riesz gases in extremely shallow confining potential, the fluctuations of the displacement of the tagged particle in one-dimensional Riesz gases have been studied in \cite{RahulRiesz} both in the long-ranged regime $0<s<1$ and effectively short-range regime $s>1$. The fluctuations of the local current in the Dyson gas ($s=0$) have also been investigated \cite{RahulDyson}. One anticipates that the tagged particle behaves diffusively when $s<-1$; the self-diffusion coefficient is unknown. 

An interesting challenge concerns large deviations parallel to those investigated in equilibrium \cite{dean06,dean08}. For instance, the Dyson gas expands diffusively, $|x|\leq 2\sqrt{Ngt}$, and the density profile is the Wigner semi-circle \eqref{Wigner}. It would be interesting to compute the probability of atypically slow expansion, $|x|\leq A\sqrt{Ngt}$ with some fixed $A<2$ during the time interval $t\in (0, T)$, and the density profile during such constrained expansion. Atypically fast expansions, $|x|\geq A\sqrt{Ngt}$ with some fixed $A>2$ during the time interval $t\in (0, T)$, are equally interesting. Alternatively, one can seek the probability that the occupied region is atypically large or small, $|x(T)|\leq A\sqrt{NgT}$ with $A\ne 2$, only at the final moment. 

\bigskip
\noindent
We are grateful to N. Oppenheimer and T. Bickel for useful correspondance. The work of KM is supported by the project RETENU ANR-20-CE40-0005-01 of the French National Research Agency (ANR). PLK is grateful to IPhT for the excellent working conditions.  

\appendix
\section{The influence of the harmonic trap}
\label{ap:trap}

Systems with particles in a confining harmonic potential are more popular (see, e.g., \cite{Forrester, Serfaty15, Petrache20, Serfaty24} and references therein) than the free systems studied in this paper. The harmonic potential naturally arises in the context of random matrices. In Coulombic systems where particles carry charges, the harmonic potential accounts for the neutralizing background. In Coulombic and other many-particle systems \cite{Lieb75, Lieb18, Lieb19, Lewin22}, an external constant background is often introduced to compensate for the repulsion between the particles.   

In this Appendix, we probe the influence of harmonic confining potential on gases with long-ranged Riesz potential. In one dimension, the density becomes stationary in the long time limit. The stationary density satisfies 
\begin{equation}
g \dashint_{-\infty}^\infty dy\,\frac{x-y}{|x-y|^{s+2}}\, \rho(y) = \epsilon x 
\end{equation}
that is solved similarly to \eqref{F1:eq} to give 
\begin{equation}
\label{rho-sol:R-inf}
\rho = \frac{\epsilon}{g}\,\frac{\cos(\pi s/2)}{\pi(s+1)}\,(r^2-x^2)^\frac{s+1}{2}  
\end{equation}
from which we deduce that the span is given by 
\begin{equation}
\label{r:R}
r = \left[\frac{Ng}{\epsilon}\,\frac{\sqrt{\pi}\,(2+s)\,\Gamma\big(1+\frac{s}{2}\big)}{\cos\!\big(\frac{\pi s}{2}\big)\,\Gamma\big(\frac{1+s}{2}\big)}\right]^\frac{1}{s+2}
\end{equation}

To determine the expansion of the one-dimensional stochastic Ries gas in the presence of the harmonic confining potential one ought to solve
\begin{equation}
\label{Riesz:1d-H}
\partial_t \rho + \partial_x\!\left[g \rho \dashint_{-\infty}^\infty dy\,\frac{x-y}{|x-y|^{s+2}}\, \rho(y,t)-\epsilon \rho x\right] = 0
\end{equation}

The similarity of the solution \eqref{scaling} and \eqref{F-sol:R}, and the stationary solution  \eqref{rho-sol:R} in the harmonic confining potential, suggests to seek the solution of \eqref{Riesz:1d-H} in the same functional form:
\begin{equation}
\label{rho:R}
\rho(x,t) = \frac{NC_s}{R^{s+2}}\,(R^2-x^2)^\frac{s+1}{2} \,, \quad C_s = \frac{\Gamma\big(2+\frac{s}{2}\big)}{\sqrt{\pi}\,\Gamma\big(\frac{3+s}{2}\big)}
\end{equation}
with amplitude fixed by normalization. Substituting \eqref{rho:R} into \eqref{Riesz:1d-H} , we find that \eqref{rho:R} provides a consistent solution if the span satisfies 
\begin{equation}
\label{R:s-eq}
\frac{dR}{dt} = \frac{gNC_s}{R^{s+1}} - \epsilon R
\end{equation}
Solving \eqref{R:s-eq} subject to $R(0)=0$ gives 
\begin{equation}
\label{R:s}
R = \left\{\frac{gN C_s}{\epsilon}\left[1-e^{-\epsilon(s+2) t}\right]\right\}^\frac{1}{s+2}  
\end{equation}
Thus, for the Riesz gas in the confining harmonic potential, the evolving density profile describing the expansion into the vacuum has the form \eqref{rho:R} throughout the evolution, with span given by \eqref{R:s}. The evolving density profile does not `feel' the confining potential when $t\ll t_c\sim \epsilon^{-1}$; the density profile approaches to the stationary density \eqref{rho-sol:R-inf}--\eqref{r:R} when $t\gg t_c$. 

In higher dimensions, $d>1$, we limit ourselves with Coulomb gases. In two dimensions, taking into account the confining harmonic potential leads to 
\begin{equation}
\label{Ginibre:H}
\partial_t F + \partial_r\!\left[\frac{g}{r}\,F\int_0^r dr'\,F(r',t)-\epsilon r F\right] = 0
\end{equation}
 generalizing Eq.~\eqref{F-Ginibre}. We anticipate that the density profile is uniform inside an expanding disk and vanishes outside the disk $r>R(t)$. Substituting $F(r,t) = r f(t)$ into \eqref{Ginibre:H},  one arrives at a closed readily solvable equation for $f(t)$. Returning to the original variables one gets
\begin{equation}
\rho(r,t) = \frac{\epsilon}{\pi g}\,\frac{1}{1-e^{-2\epsilon t}} 
\end{equation}
Thus, the density is again uniform inside the disk of radius $R(t)$. The radius of the disk is fixed by mass conservation:
\begin{equation}
R(t) = \sqrt{\frac{gN}{\epsilon}\,\big(1-e^{-2\epsilon t}\big)}
\end{equation}
Again, we observe an exponential convergence to the stationary case, with a crossover time of order $1/\epsilon$.

Similarly for the $d-$dimensional Coulomb gas in the harmonic confining potential $\epsilon r^2/2$, the radius of the occupied ball increases according to 
\begin{equation}
\label{RH:Coulomb}
R(t) = \left[\frac{gN}{\epsilon}\,\big(1-e^{- d \epsilon t}\big)\right]^{1/d}
\end{equation}
and the density inside the ball is uniform and given by
\begin{equation}
\label{Friedmann:H}
\rho(r,t) = \frac{\epsilon d}{\Omega_d g}\,\frac{1}{1-e^{- d \epsilon t}} 
\end{equation}
For the free Coulomb gas ($\epsilon=0$), Eqs.~\eqref{RH:Coulomb}--\eqref{Friedmann:H} reduce to Eqs.~\eqref{R:Coulomb}--\eqref{Friedmann}. 

We emphasize that for the Coulomb gas in the harmonic potential, the problem remains well-defined when the number of particles is infinite, $N=\infty$. The density is again uniform in the entire space $\mathbb{R}^d$. In the earlier time regime, $t\ll \epsilon^{-1}$, the density decays according to \eqref{Friedmann}, then it saturates to the finite value $\rho\to \rho_\infty=\frac{\epsilon d}{\Omega_d g}$, when $t\gg \epsilon^{-1}$. In the stochastic Coulomb gas with infinitely many particles adding the confining harmonic potential is akin to adding the cosmological constant. The signs are reversed: the interactions are repulsive (in contrast with the attractive gravitational interactions), and the `cosmological constant'  ($-\epsilon$) is negative rather than positive. In the deterministic framework, there is a similar analogy of the Coulomb explosion with cosmic expansion in the presence of the negative cosmological constant  \cite{Kolomeisky19,Kolomeisky20,Kolomeisky21,cosmic}.

\bibliography{vacuum}

\end{document}